\documentclass[twocolumn,prl,10pt,aps,floatfix]{revtex4-1}
\usepackage{graphicx}
\usepackage[colorlinks,linkcolor=blue,citecolor=blue,urlcolor=blue]{hyperref}
\usepackage{color}

\begin{document}

\title{High temperature dynamics in quantum compass models}
\author{A.K.R. Briffa$^{1}$}
\author{X. Zotos$^{1,2,3,4}$}
\affiliation{$^1$ITCP and CCQCN, Department of Physics, 
University of Crete, 71003 Heraklion, Greece}
\affiliation{$^2$Foundation for Research and Technology - Hellas, 71110 
Heraklion, Greece}
\affiliation{$^3$ Max-Planck-Institut f\"ur Physik Komplexer Systeme,
01187 Dresden, Germany}
\affiliation{
$^4$Leibniz Institute for Solid State and Materials Research IFW Dresden, 
01171 Dresden, Germany} 

\date{\today}

\begin{abstract}
We analyze the high temperature spin dynamics of quantum compass models 
using a moment expansion. We point out that the evaluation of moments maps to 
the enumeration of paths in a branching process on the lattice. This mapping 
to a statistical mechanics combinatorics problem provides an elegant 
visualization of this analysis. We present results for the time dependent 
spin correlation function (which is of relevance to NMR experiments) for two 
compass models: the honeycomb (- Kitaev) and 2D compass model.
Furthermore, following this novel approach, we argue that in 
quantum compass models the spin correlations are generically strongly 
anisotropic and short ranged.
\end{abstract}
\maketitle

\section{Introduction}
Over the years quantum Compass models have proved especially relevant 
in the modeling 
of materials with multi-orbital degrees of freedom\cite{komskii}; 
these Hamiltonians 
describe fictitious spin degrees of freedom that have fully anisotropic 
spatial interactions\cite{brink}. Currently they are very prominent in both 
theoretical and experimental studies of the physics of 
iridium-oxide materials\cite{iridium}, 
the $\alpha-$RuCl$_3$ compound\cite{knolle,vojta} and finally the purported 
application of the Kitaev model\cite{kitaev} to quantum computing. Neutron 
scattering\cite{neutron} and NMR experiments\cite{nmr} 
have provided convincing 
evidence that Kitaev model physics dominates the effective Hamiltonian 
describing 
$\alpha-$RuCl$_3$, although the inclusion of further interactions, 
especially Heisenberg 
exchange, seem to be necessary for a consistent description of the experiments. 
Additionally, thermal conductivity experiments could shed more light 
on the relevant 
interactions in these exotic quantum magnets.

The canonical theoretical approach to the spin dynamics analysis 
of the Kitaev model is the elegant and powerful Majorana fermion 
solution\cite{kitaev}. So far most studies have focused on the zero 
temperature 
limit\cite{knolle,vojta}. However, at finite temperatures Monte Carlo 
methods have been used\cite{nasu} to sample the auxiliary Majorana fields 
and there are Exact Diagonalization studies 
on restricted lattices\cite{metavitsiadis}.

In the high temperature limit, a long standing approach to spin dynamics 
has been the moment method\cite{muller}. 
In this work, we point out that the peculiar 
structure of compass model Hamiltonians maps the moment enumeration to a
branching model, though 
there are certain extra initial constraints specific to each model.
In the following, we discuss as examples the compass model on the 
two dimensional honeycomb lattice (the so called Kitaev model) 
and the two-dimensional compass model on a square lattice. 

Our main focus is to obtain the salient features 
of the exact spin autocorrelation functions from the mapping to
model branching processes and compare them to results obtained 
by diagonalization of the Hamiltonian (ED) on finite size lattices. 
The validity of the mapping is supported by the agreement 
with frequency spectra obtained by ED. It is amusing that such simple 
combinatorial models reproduce the essential features of these quantum 
many-body models, providing another instance of correspondence between 
quantum mechanical and statistical mechanics problems.
It would be interesting also to consider the inverse procedure, where 
a classical statistical mechanics problem can be solved by a corresponding 
quantum compass models. 

\section{Method}
Primarily we are interested in the evaluation of the spin 
autocorrelation function 
\begin{equation}
C(t)=< \tau_0^z(t) \tau_0^z>
\end{equation}
\noindent
where $\tau^{\alpha},~~\alpha=x,y,z$ are Pauli spin-1/2 operators,
and $\tau_0^z(t)=e^{+iHt}\tau_0^z e^{-iHt}$, $\hbar=1$.
We will also briefly discuss further spatial correlations,
\begin{equation}
C_{ij}(t)=< \tau_j^z(t) \tau_i^z>,~~~i \ne j.
\end{equation}

\noindent
$< ... >$ denotes a thermal average at temperature $T$ (with $\beta=1/T$).
In the infinite temperature limit, $\beta \rightarrow 0$, the autocorrelation 
function reduces to a trace over all the Hilbert space,
\begin{equation}
C(t)=\frac{1}{2^L} {\textrm tr}~\tau_0^z(t) \tau_0^z
\end{equation}
\noindent
$L$ being the number of spins on the lattice.

Expanding in powers of time,
\begin{equation}
C(t)=\sum_{k=0}^{\infty} \frac{(-1)^k}{(2k)!} \mu_{2k} t^{2k}
\label{auto}
\end{equation}
\noindent
the autocorrelation function analysis reduces to the evaluation of the 
moments $\mu_{2k}$\cite{muller},

\begin{equation}
\mu_{2k}=\frac{1}{2^L} {\textrm tr}~\tau_0^z {\cal L}^{2k} \tau_0^z 
\label{liouville}
\end{equation}

\noindent
where ${\cal L} =[H,A]=HA-AH$ is the Liouville operator.
The time Fourier transform of $C(t)$,

\begin{equation}
S(\omega)=\int_{-\infty}^{+\infty} C(t) e^{+i\omega t} dt
\end{equation}

\noindent
can alternatively be evaluated by an extension in complex frequencies $z$,
\begin{equation}
c(z)=\int_0^{+\infty} C(t) e^{-z t} dt,~~\Re(z) > 0, 
\end{equation}

\begin{equation}
S(\omega)=\lim_{\eta \rightarrow 0^+} 2\Re [c(\eta -i \omega)]. 
\end{equation}

\noindent
$c(z)$ is then conveniently expressed as a continued fraction 
expansion,  

\begin{equation}
c(z)=\frac{1}{z+\frac{\Delta_1}{z+\frac{\Delta_2}{z+..}}}
\end{equation}

\noindent
with the coefficients $\Delta_n$ related the moments $\mu_{2k}$ by 
recursion relations\cite{muller}.

In the following we will present an approximate evaluation of the 
moments $\mu_{2k}$ and the corresponding structure of $\Delta_n$ 
by mapping them to combinatorial branching models.

\section{Kitaev model}
\begin{figure}[!ht]
\begin{center}
\includegraphics[width=1.2\linewidth, angle=0]{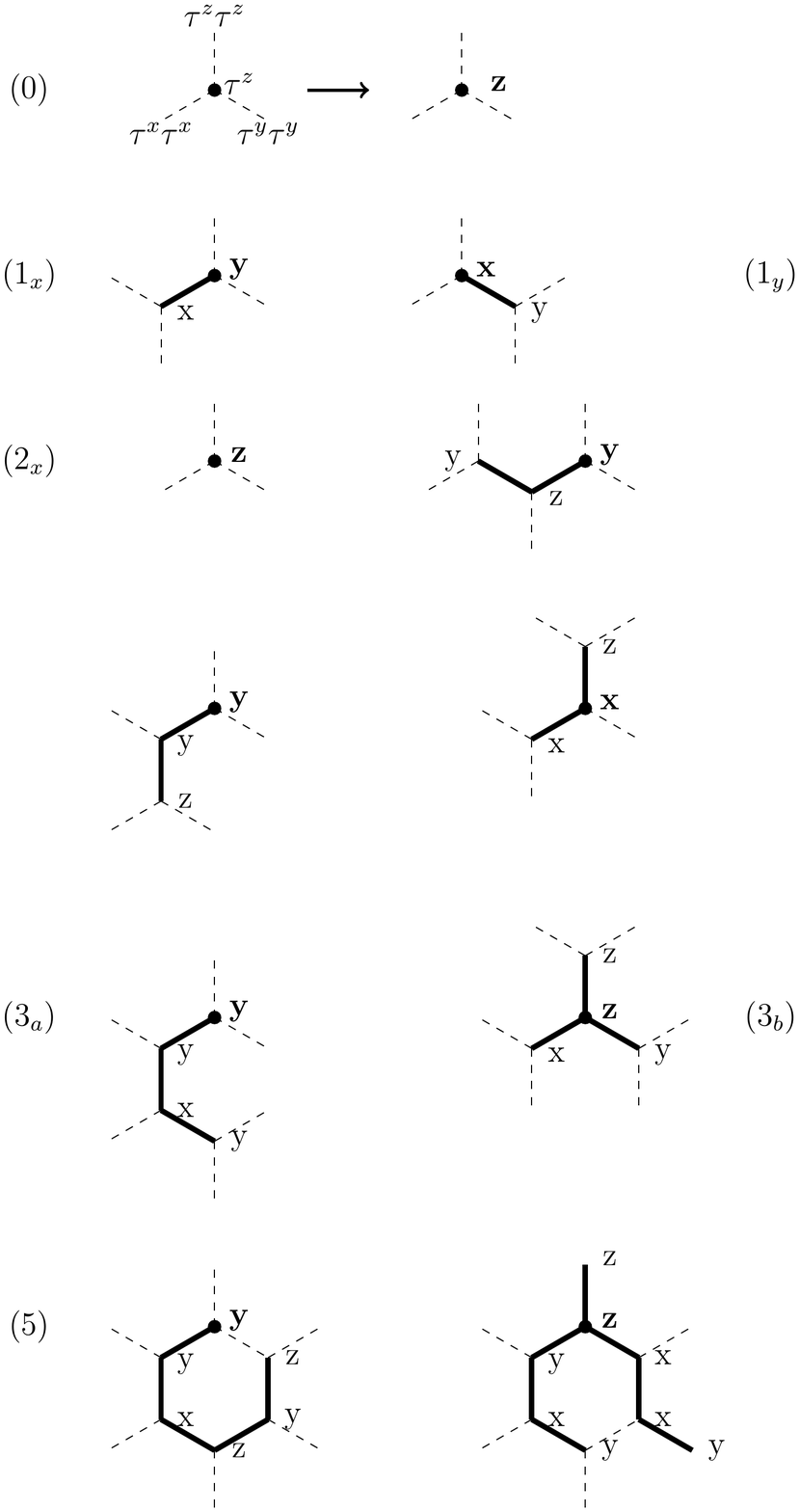}
\caption{Indicative branching processes on the Kitaev model 
on a honeycomb lattice.}
\label{ktv}
\end{center}
\end{figure}

The Kitaev model on a honeycomb lattice is given by the Hamiltonian,
\begin{equation}
H=-J_x\sum_{<ij>_x} \tau_i^x\tau_j^x
  -J_y\sum_{<ij>_y} \tau_i^y\tau_j^y
  -J_z\sum_{<ij>_z} \tau_i^z\tau_j^z
\end{equation}

\noindent 
where $<ij>_{x,y,z}$ denotes the nearest neighbor bonds in the three directions 
on the lattice, with the convention indicated in Fig.\ref{ktv}(0).
The central operator $\tau^z$ (that subsequently is abbreviated 
just by $\bf z$) represents the operator $\tau_0^z(t)$ at 
the initial time $t=0$. 

We find that repeatedly applying the Liouville operator $\cal L$, as in 
(\ref{liouville}), creates strings of operators depicted in the subsequent 
parts of Fig.\ref{ktv}. 
Because of the 3-fold geometry of the honeycomb lattice, 
the string operators are organized into three branches, each one starting 
with an $xx$, $yy$ or $zz$ bond. 

Fig.\ref{ktv}$(1_x), (1_y)$ 
indicate the two possible operator strings
created by one application of $\cal L$ on  $\tau_0^z(t=0)$. The third sting, 
along the z-bond, vanishes at this order. In ($2_x$)  the four possible (non-vanishing) 
strings created by the application of $\cal L$ on ($1_x$) are shown (note that 
there are an additional four, symmetrically related, strings for ($1_y$) which 
are not shown). Then, the application of the trace in (\ref{liouville}) gives zero 
for all but the first diagram shown in Fig.\ref{ktv}$(2_x)$: the zero-length string 
which is just the operator $\tau_0^z$. Consequently there are only two contributions 
to the second moment, $\mu_2$: this one and the analogous diagram arising from the 
application of $\cal L$ on ($1_y$).The problem of applying $\cal L$ 
on a operator string 
is therefore mapped to a branching process in three directions, where at each iteration 
the tail operator in a branch either disappears or a new one is created in one of the 
three possible directions (although with some specific restrictions at the origin which 
we will discus later).

In the high temperature limit, the trace of any string of non-zero length vanishes. 
Thus, evaluating the moments reduces to counting all of the possible branching processes 
that culminate in the complete annihilation of the strings (i.e. they return to just a 
single $\tau_0^z$ operator). Each branching configuration contributes a factor of 
$(\pm 2iJ_\alpha)$ from each application of $\cal L$ to the value of the contribution to 
a moment. This arises from the spin commutation involving the bond-$\alpha$; the factor for 
annihilating a particular link turns out to always have the opposite sign 
to that corresponding 
to its creation. As an example, the third order diagram shown in Fig.\ref{ktv}$(3a)$ 
represents the operator string: $(-2iJ_y)(-2iJ_z)(+2iJ_x)\tau_3^y\tau_2^x\tau_1^y\tau_0^y$
Henceforth we will only consider $J_x=J_y=J_z$, although generalising to the 
anisotropic case is more complex but feasible. 
In summary, it therefore seems that the minimal model is branching in 
three independent directions. However, as we will now discuss, this is not quite adequate. 
There are some extra initial constraints and also some "higher order" processes that 
need consideration. 

We find that the branching model obeys the following {\it rules}:

(i) side-branching is not possible and the operator 
at a vertex is given by the direction of the missing bond;

(ii) applying $\cal L$ in the middle of a branch annihilates 
the entire operator string; 

(iii) independent branching along all three directions is not possible: 
there are initial restrictions. 
A branch along the $z$-direction cannot be created at 
first order. Instead, it can only be initiated 
when either an $x$-direction or a $y$-direction branch already exists, 
as depicted in Fig.\ref{ktv}($2_x$). This 
initial restriction effects the lowest order moments 
and thus the high frequency behavior of the spectral function. 

(iv) Branches must be annihilated in the reverse order to which 
they were created or the 
contribution from the entire string vanishes. 

(v) Possible higher order processes are not taken into account where 
the branching 
structure is modified e.g. at the origin and then eventually reconstructed and 
annihilated.

(vi) Finally, as shown in Fig.\ref{ktv}$(5)$, the branching process 
is self-avoiding.

\begin{figure}[!ht]
\begin{center}
\includegraphics[width=1.0\linewidth, angle=0]{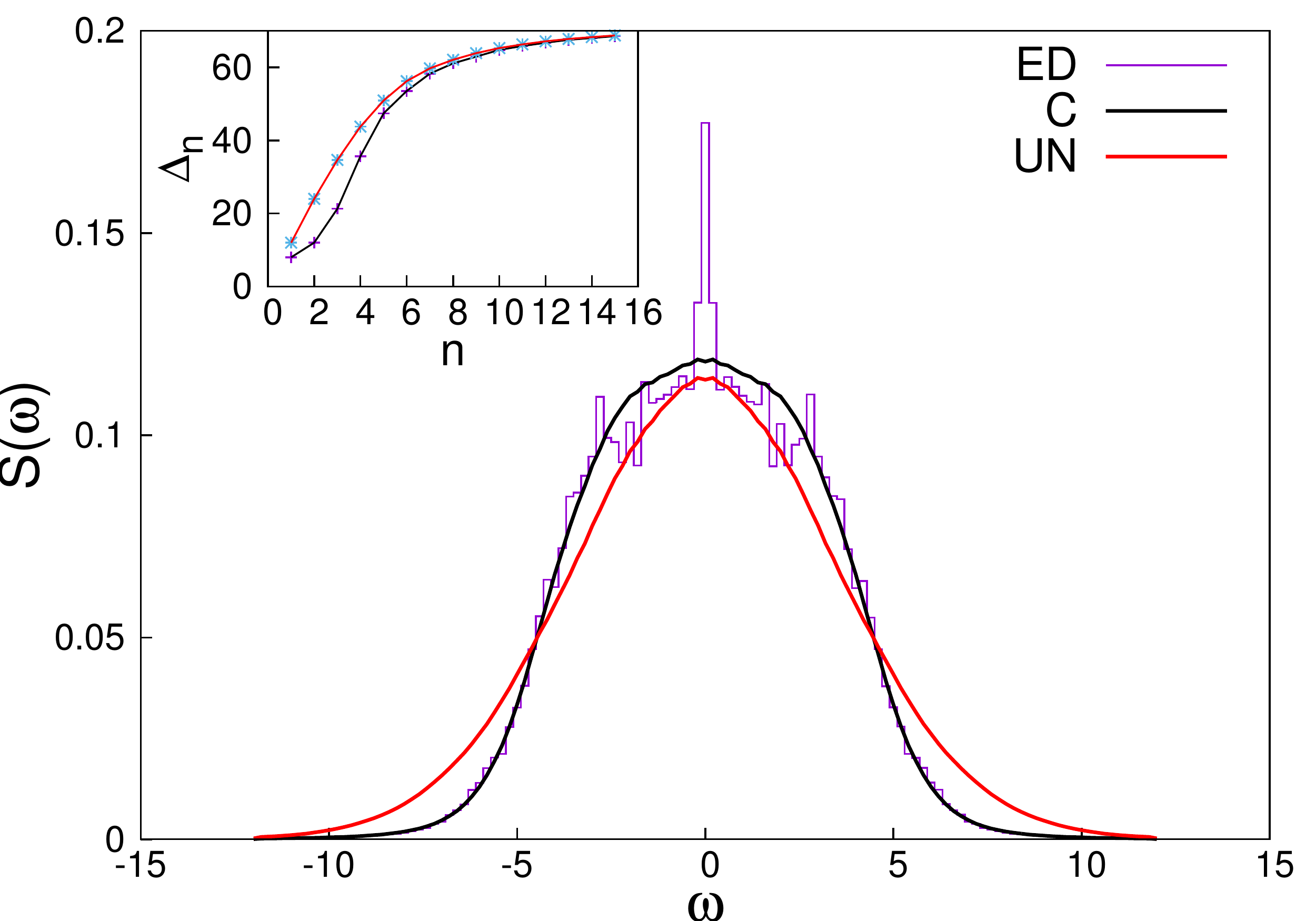}
\caption{Spectrum of Kitaev model, by ED, branching model without (UN) 
and with (C) initial constraints; inset: corresponding $\Delta_n$ values.}
\end{center}
\label{ktvspectrum}
\end{figure}

In the inset of Fig.\ref{ktv} 
we show the values of $\Delta_n$ evaluated from the 
moments $\mu_{2k}$ for the two models, the branching process without (shown in 
red) and with (shown in black) the initial constraints. It is clear that for 
large values of $n$ the $\Delta_n$'s asymptotically coincide. In these 
branching model calculations, however, we have omitted the self-avoidance constraint.


To estimate the applicability of the branching model we compare in 
Fig.\ref{ktv} the spectra obrained by the branching models with (C)
and without (UN) initial constraints. Also shown are 
results from an exact diagonalization study (ED) on a lattice of 4 by 6 spins. 
with periodic boundary conditions. Here we calculated in the high temperature limit
using the microcanonical Lanczos method\cite{micro}.
100 Lanczos iterations were used to converge to an infinite temperature state and 
400 further iterations to obtain the continuous fraction expansion.
The $\delta-$function peak at zero frequency is a finite size effect due 
to an excess of degenerate states in small lattices. 
The agreement is fair with the model when including initial constraints (despite 
omitting the self-avoidance requirement) and can be fitted by a form 
$S_{fit}(\omega) \sim a/(b+\omega^6)$.
We should stress that this frequency dependence is way off a Gaussian 
or Lorentzian form.

\section{2D compass model}

\begin{figure}[!ht]
\begin{center}
\includegraphics[width=1.0\linewidth, angle=0]{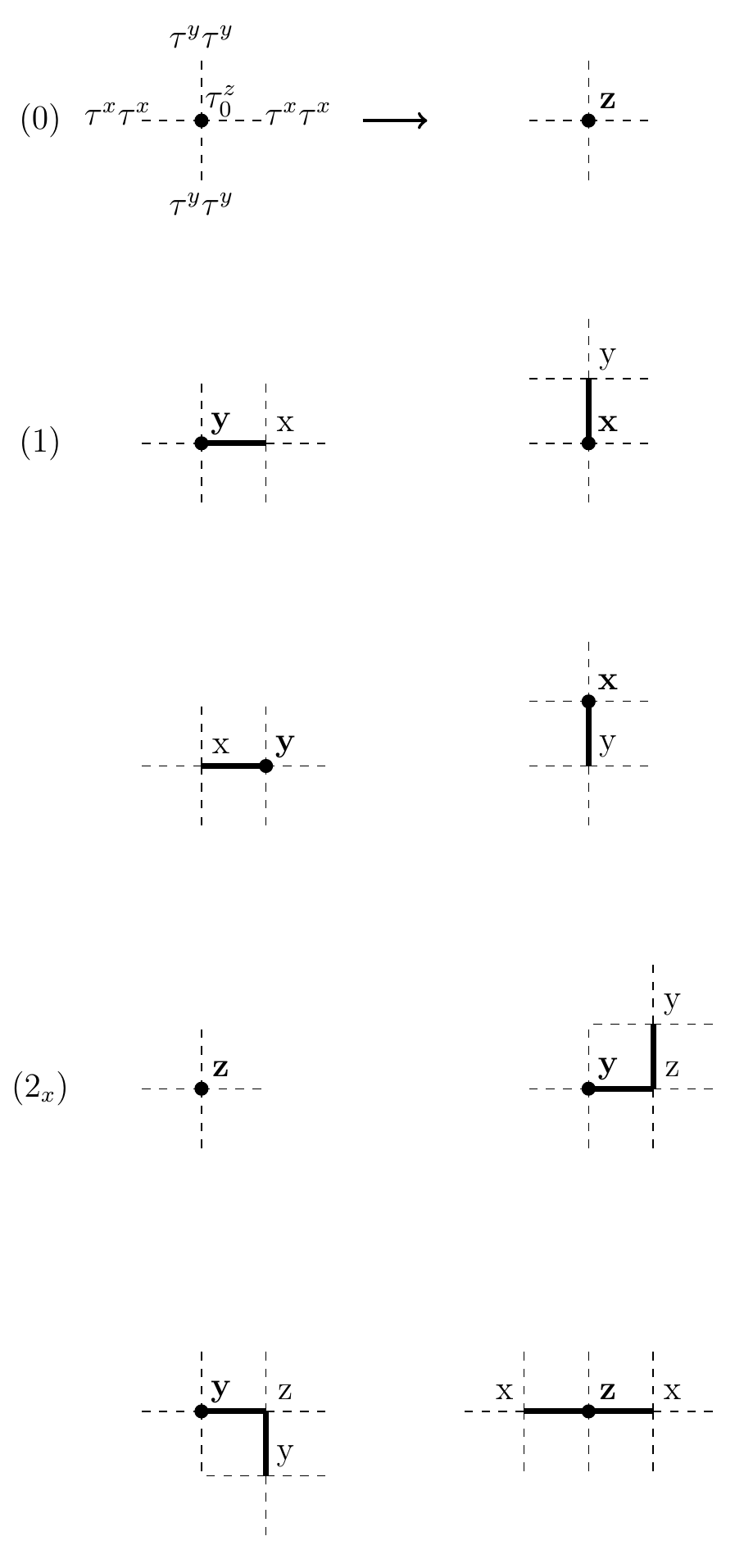}
\caption{Indicative branching processes for the 2D compass model 
on a square lattice.}
\label{compass}
\end{center}
\end{figure}

Next, we will study the 2D-compass model, using the same approach as previously. 
The Hamiltonian is given by,
\begin{equation}
H=-J_x\sum_{<ij>_x} \tau_i^x\tau_j^x
  -J_y\sum_{<ij>_y} \tau_i^y\tau_j^y
\end{equation}

\noindent
with the bond-labeling convention used depicted in Fig.\ref{compass}(0).

Analogously to our discussion of the Kitaev model, repeated applications of the 
Liouville operator creates strings of operators which can propagate with up to four 
branches: see Fig.\ref{compass}. The same procedure for obtaining 
the moments $\mu_{2k}$ and corresponding $\Delta_n$ of this combinatorial model can 
then be applied, but with 
the same reservations on possibly omitted high-order processes.

This time there are a different set of restrictions specific to the 2D 
compass model. 
Firstly, we find that the operator strings must propagate alternatively 
along $x$-links 
and $y$-links. Trying to extend the string along two consecutive $x$-links 
(or $y$-links) 
causes the entire string to vanish. This also means that the internal spin 
operators along 
the length of the string are all $\tau^z$'s. Secondly, we note that, 
as there was in the 
Kitaev model, there are initial constraints on the order in which the four 
different branches 
can be created and subsequently destroyed. Finally, the branching model is 
again self-avoiding. 

\begin{figure}[!ht]
\begin{center}
\includegraphics[width=1.0\linewidth, angle=0]{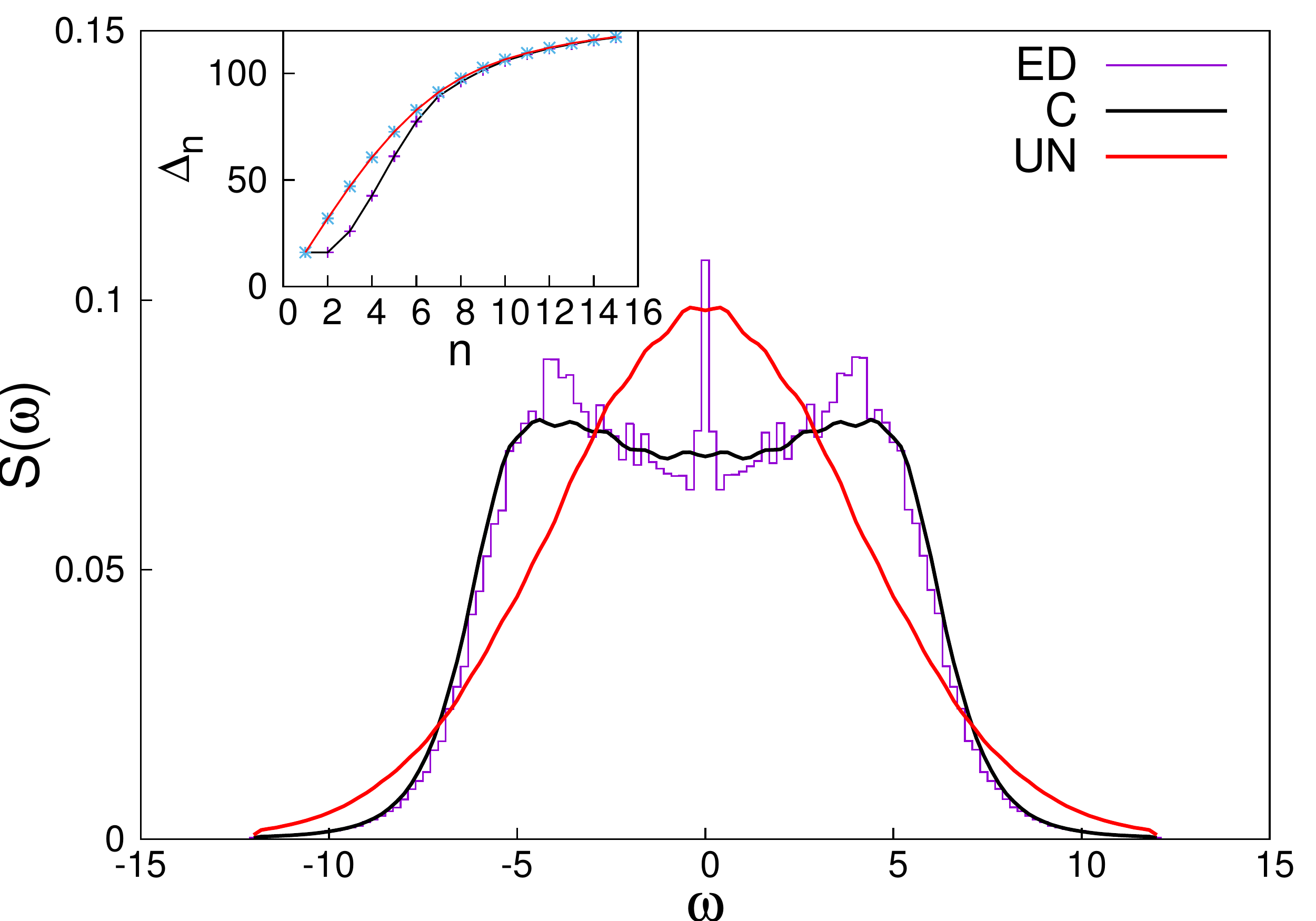}
\caption{Spectrum of the 2D compass model by ED, branching model without (UN) 
and with (C) initial constraints; inset: corresponding $\Delta_n$ values.}
\label{compassspectrum}
\end{center}
\end{figure}

To investigate the infinite temperature spin-autocorrelation for the isotropic
2D compass model we have calculated a finite number of moments for two 
different branching models: one which includes the initial constraints, and 
one with out. In both cases, however, we have again omitted the self-avoidance 
requirement. The $\Delta_n$ parameters that we have calculated are shown in the 
inset of Fig.\ref{compassspectrum}. 

The subsequent $S(\omega)$ spectrum is shown in Fig.\ref{compassspectrum} 
for the branching model including the initial branch-creation-order 
constraint (C) and without this initial constraint (UN). These are 
compared to the spectrum obtained by a microcanonical Lanczos calculation (ED) 
on a lattice of 4 by 6 spins. It is interesting to 
see that even the non-monotonic behaviour of $S(\omega)$ is reproduced 
when including the initial branch--creation-order constraint. In contrast, 
the unconstrained branching behaviour cannot capture this, indicating that 
it arises due to the degeneracy of the first two $\Delta$ values. 


By using the same approach of expanding the time dependance as a function 
of the Liouville operator, we can also consider further-range spin 
correlation in the high temperature limit. 
\begin{equation}
C_{0j}(t)=\frac{1}{2^L} {\textrm tr}~\tau_j^{\alpha}(t) \tau_0^{\alpha}.
\end{equation}
\noindent
Consequently, we deduce, that for $j\ne 0$ all 
correlations vanish in the infinite temperature limit.
This occurs because the initial 
pair of operators $\tau_j^{\alpha}\tau_0^{\alpha}$ cannot
be reduced to the identity by application of $\cal L$.
We can also consider finite temperatures by expanding 
$e^{-\beta H}$ in powers of $H$. 
Then we can plausibly argue that we obtain a finite value 
for the correlations $<\tau_j^{\alpha}(t)\tau_0^{\alpha}>$, 
only when $j$ is a nearest neighbor site in the $\alpha-$ lattice direction.
However, further-neighbour correlations all vanish identically at all 
temperatures. Again, this is because it is impossible to reduce 
any of the resulting operator strings to the identity by application of
either $\cal L$ or $H$. This result has already been found for the 
Kitaev model using the Majorana fermion approach\cite{baskaran}.
Following the same line of reasoning, in the 2D compass model 
the $\tau_j^{\alpha}\tau_0^{\alpha}$ 
correlations are nonzero only at finite temperatures and only when the spins 
$\tau_0^{\alpha},~\tau_j^{\alpha}$ are on the same $\alpha-$ direction chain. 
Similar considerations also apply to higher dimensionality quantum 
compass models. 

In conclusion, the above analysis, to an extent heuristic, 
provides an interesting novel perspective. Firstly, we have 
discovered that the high temperature spin-autocorrelation 
function of quantum compass models can be mapped onto a statistical-mechanical 
branching problem (along with some additional, model-specific restrictions). 
We have discussed the results from calculating a finite number of moments 
for two different isotropic compass models, investigating the effects of 
appropriately 
including the model-specific initial restricts on the branching-scheme. 
Despite omitting 
the self-avoidance requirement, this provided a useful insight into the shape of these 
correlation functions. These techniques also have the potential to be generalised 
to anisotropic couplings $J_x\ne J_y \ne J_z$ and the evaluation of finite temperature 
correlations. It may be feasible to obtain analytic expressions for the moments of a 
constraint-free branching model, but it is clear that an exact evaluation of the 
autocorrelation function requires the implementation of self-avoidance, which 
is well-known to be a highly non-trivial problem. 
This raises the interesting perspective of whether the 
inverse procedure could be used to solve self-avoiding statistical mechanics 
problems by means of a mapping onto a corresponding quantum compass model.
It appears to be another instance of correspondance between 
quantum mechanical and statistical mechanics problems. Finally, we note  
that other types of correlations functions, 
for example those related to transport 
properties, can also be studied using this technique. 

\section{Acknowledgments}

This work was supported by the European Union Program
No. FP7-REGPOT-2012-2013-1 under Grant No.~316165.
A.K.R.B. acknowledges the hospitality of MPI - PKS extended to her 
during visits. X.Z. acknowledges fruitful discussions with B. B\"uchner, 
C. Hess, S. Miyashita, H. Tsunetsugu, the hospitality of 
the Institute for Solid State Physics - U. Tokyo and
support by the Alexander von Humboldt Foundation.

\end{document}